\documentclass[conference]{IEEEtran}

\usepackage[cmex10]{amsmath}
\usepackage{amssymb}

\usepackage{cases}
\usepackage{multirow}
\usepackage{empheq}

\usepackage{cite}
\usepackage{verbatim}

\ifCLASSINFOpdf
  \usepackage[pdftex]{graphicx}
\else
  \usepackage[dvips]{graphicx}
\fi
\usepackage[caption=false,font=footnotesize]{subfig}

\usepackage{color}
\usepackage{multirow}

\begin{document}

\sloppy

\title{Writing on Dirty Flash Memory}

\author{
  \IEEEauthorblockN{Yongjune Kim and B. V. K. Vijaya Kumar}
  \IEEEauthorblockA{Department of Electrical \& Computer Engineering, Data Storage Systems Center (DSSC)\\
    Carnegie Mellon University\\
    Pittsburgh, PA, USA\\
    Email: yongjunekim@cmu.edu, kumar@ece.cmu.edu}
}


\maketitle

\begin{abstract}
The most important challenge in the scaling down of flash memory is its increased inter-cell interference (ICI). If side information about ICI is known to the encoder, the flash memory channel can be viewed as similar to Costa's ``writing on dirty paper (dirty paper coding).'' We first explain why flash memories are \emph{dirty} due to ICI. We then show that ``dirty flash memory'' can be changed into ``memory with defective cells'' model by using only one pre-read operation. The asymmetry between write and erase operations in flash memory plays an important role in this change. Based on the ``memory with defective cells'' model, we show that additive encoding can significantly improve the probability of decoding failure by using the side information.
\end{abstract}

\section{Introduction}

Aggressive scaling down of memory cell size has driven the continuous growth of flash memory density. However, the scaling down leads to many challenges. One primary challenge is the increased inter-cell interference (ICI) between adjacent (neighboring) flash memory cells \cite{Prall2007}. As the distance between adjacent cells decreases due to scaling down, flash memory cells suffer from higher ICI \cite{Prall2007, Lee2002}.


In order to cope with the ICI, various approaches have been proposed. Device level approaches such as new materials and novel cell structures try to reduce the parasitic capacitances between adjacent cells \cite{Prall2010}. At circuit and architecture levels, several write (program) schemes and all bitline (ABL) architecture were proposed to deal with the ICI \cite{Park2008zeroing, Shibata2008, Li2009}.

Recently, strong error control codes (ECC) such as low-density parity check (LDPC) codes and signal processing have been also investigated \cite{Dong2011soft, Wang2011soft, Dong2010compensation}. The disadvantage of ECC with soft decision decoding and signal processing is the degradation of read speed due to multiple reads needed to obtain the soft decision values. In addition, modulation coding has been investigated to reduce some data patterns which are vulnerable to ICI \cite{Berman2011constrained, Kim2013modulation}. The significant redundancy of modulation coding is an important drawback.

In this paper, we propose a scheme that uses the side information corresponding to the ICI. In particular, the encoder uses this side information to improve the decoding failure probability, but at the expense of decreased write speed. The decrease in write speed may be acceptable for memory systems, since the write operation is typically not on the critical path because of write buffers available in the memory hierarchy~\cite{Jagmohan2010a, Hennessy2002}. In addition, the read speed degradation would be more critical in applications such as one-time programmable (OTP) flash memories.

Theoretically, the proposed scheme can be explained by Gelfand-Pinsker problem. The Gelfand-Pinsker problem assumes that only the encoder knows noncausally the side information of the channel~\cite{Gelfand1980}. There are two famous examples of Gelfand-Pinsker problem: ``Writing on dirty paper (dirty paper coding)'' in \emph{communication} and ``memory with defective cells'' in \emph{storage}~\cite{Costa1983dpc, Kuznetsov1974, ElGamal2011}.

\begin{figure}[t]
   \centering
   \includegraphics[width=0.45\textwidth]{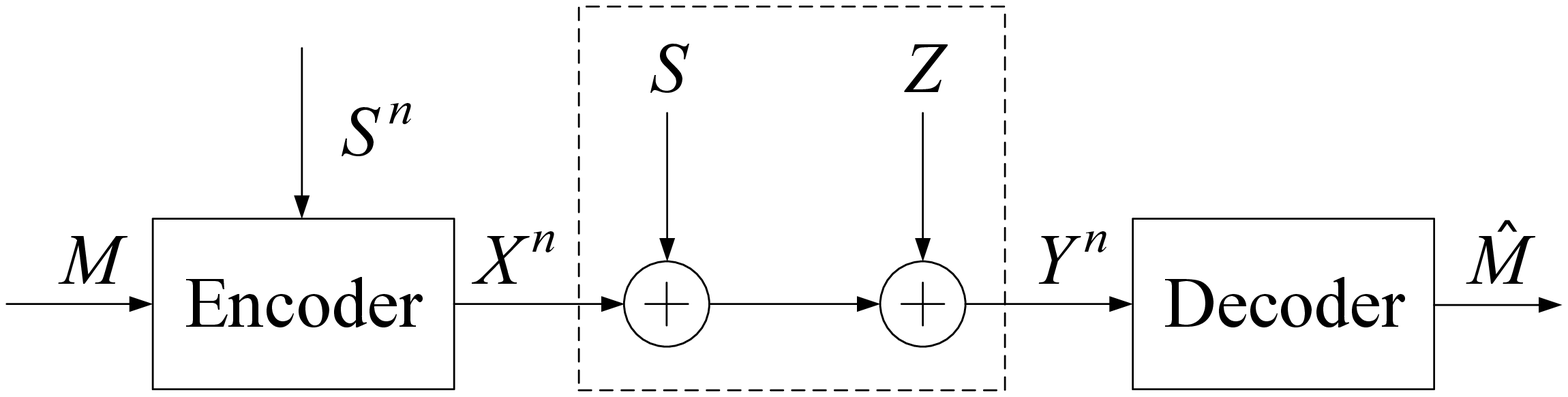}
   \caption{Dirty paper channel where $M$ and $\hat{M}$ represent a message and its estimate, respectively \cite{ElGamal2011}.}
   \label{fig:channel_dpc}
   \vspace{-5mm}
\end{figure}

Costa's writing on dirty paper considers the following channel~\cite{Costa1983dpc}:
\begin{equation}\label{eq:costa}
Y = X + S + Z
\end{equation}
where $X$ and $Y$ are the channel input and output, respectively. Also, $S \sim N\left(0, \sigma_S^2\right)$ is an interference and $Z \sim N\left(0, \sigma_Z^2\right)$ is an additive noise. Assume that the channel input satisfies an average power constraint $\frac{1}{n} \sum_{i=1}^{n}{X_i^2} \le P$ for the channel input vector $X^n = \left(X_1, \cdots, X_n \right)$. If neither the encoder nor the decoder knows $S$, the capacity is given by
\begin{equation}\label{eq:capacity_dpc_min}
C_{\textrm{min}} = \frac{1}{2} \log_2\left(1 + \frac{P}{\sigma_S^2 + \sigma_Z^2}\right).
\end{equation}
As shown in Fig.~\ref{fig:channel_dpc}, if the encoder knows the entire interference vector $S^n = \left(S_1, \cdots, S_n \right)$ prior to transmission, the capacity is given by
\begin{equation}\label{eq:capacity_dpc_max}
C_{\textrm{max}} = \frac{1}{2} \log_2\left(1 + \frac{P}{\sigma_Z^2} \right)
\end{equation}
where the effect of the interference $S$ is completely cancelled out \cite{Costa1983dpc}.

\begin{figure}[t]
   \centering
   \includegraphics[width=2.4in]{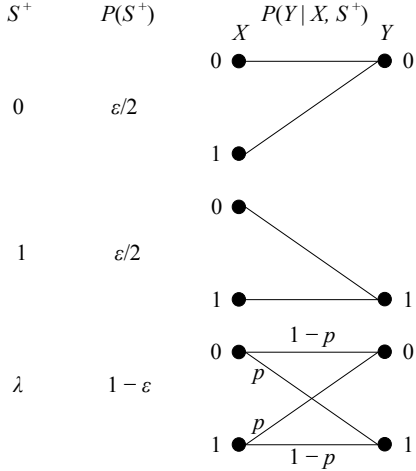}
   \vspace{-3mm}
   \caption{Memory with defective cells.}
   \label{fig:channel_defect}
   \vspace{-5mm}
\end{figure}

The memory with defective cells was introduced by Kuznetsov and Tsybakov~\cite{Kuznetsov1974}. A binary memory cell is called defective if its cell value is stuck-at a particular value regardless of the channel input. As shown in Fig.~\ref{fig:channel_defect}, this channel model has a ternary defect information $S^+ \in \{0, 1, \lambda\}$ whereas the channel input $X$ and output $Y$ are binary. The state $S^+=0$ corresponds to a stuck-at 0 defect that always outputs a 0 independent of its input value, the state $S^+=1$ corresponds to a stuck-at 1 defect that always outputs a 1, and the state $S^+=\lambda$ corresponds to a normal cell that can be modelled by a binary symmetric channel (BSC) with crossover probability $p$. The probabilities of these states are $\varepsilon / 2$, $\varepsilon / 2$ (assuming a symmetric defect probability), and $1 - \varepsilon$, respectively~\cite{Heegard1983capacity}.

If neither the encoder nor the decoder knows $S^+$, the capacity is given by
\begin{equation} \label{eq:capacity_bdc_min}
C_{\textrm{min}}^{+} = 1 - h \left(\left(1 - \varepsilon\right)p + \frac{\varepsilon}{2}\right)
\end{equation}
where $h\left( x\right) = -x \log_2 x - \left(1-x\right) \log_2\left(1-x\right)$. Note that \eqref{eq:capacity_bdc_min} equals the capacity of a BSC with crossover probability $\widetilde{p} = \left(1 - \varepsilon \right) p + \frac{\varepsilon}{2}$. If the encoder or the decoder knows $\left(S^{+}\right)^n = \left(S_1^+,\cdots,S_n^+\right)$ of $n$ memory cells, the maximum capacity of memory with defective cells can be achieved \cite{Heegard1983capacity}. The capacity is given by
\begin{equation}\label{eq:capacity_bdc_max}
C_{\max}^{+} = \left(1 - \varepsilon \right) \left( 1 - h \left(p \right) \right).
\end{equation}

The common ground of writing on dirty paper and memory with defective cells is that in both these cases only the encoder knows $S^n$ or $\left(S^{+}\right)^n$. Thus, these two examples can be categorized as Gelfand-Pinsker problem.

Suppose that $S^n$ represents the ICI of $n$ flash memory cells. Then, the important question is how the encoder knows $S^n$ accurately. Unfortunately, it is difficult for the encoder to know $S^n$ due to flash memory's properties (the details are explained in Section~\ref{subsec:fmc}). Thus, we change flash memory channel with the ICI into flash memory with defective cells, which means that the encoder uses the side information of defects $\left(S^+\right)^n$ rather than the side information of ICI $S^n$. The asymmetry between write and erase operations of flash memory plays a pivotal role in this change (the asymmetry between write and erase operations is explained in Section~\ref{subsec:basic}). It is worth mentioning that two famous examples of Gelfand-Pinsker problem come together in flash memories.


After changing the flash memory channel with the ICI into the model of memory with defective cells, we consider the additive encoding to use the side information $\left(S^+\right)^n$ \cite{Tsybakov1975, Heegard1983capacity}. The numerical results show that the proposed scheme can improve the probability of decoding failure significantly due to $\left(S^+\right)^n$ available at the encoder. 

Recently, coding schemes that use the side information at the encoder have drawn attention for phase change memories (PCM) and write once memory (WOM) codes~\cite{Jagmohan2010a, Kurkoski2013}. We will focus on the ICI of flash memories and propose a scheme to obtain the side information corresponding to the ICI efficiently and improve the probability of decoding failure by using this information at the encoder.

The rest of this paper is organized as follows. Section~\ref{sec:background} reviews the basics of flash memory such as flash memory structure, asymmetry between write and erase operations, and ICI. Section~\ref{sec:dirtyflash} explains why flash memories are dirty and how it can be formulated into the model of memory with defective cells. Also, the additive encoding for this model will be briefly explained. After showing the numerical results in Section~\ref{sec:result}, Section~\ref{sec:conclusion} concludes the paper.

\section{Background}\label{sec:background}

\subsection{Flash Memory Basics} \label{subsec:basic}

\begin{figure}[t]
\centering
\includegraphics[width=2.3in]{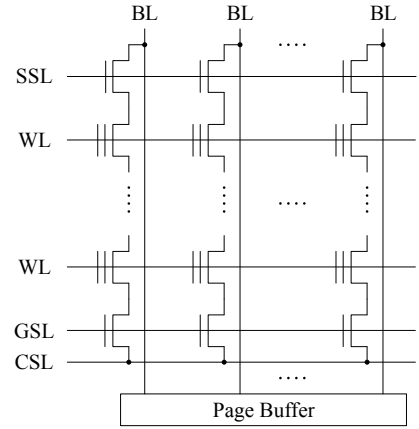}
\vspace{-2mm}
\caption{Flash memory block where SSL, GSL, and CSL denote string select line, ground select line, and common source line, respectively \cite{Suh1995}.}
\label{fig:structure}
\vspace{-5mm}
\end{figure}

Each flash memory cell is a floating gate transistor whose threshold voltage can be configured by controlling the amount of electron charges in the floating gate \cite{Dong2010compensation}. More electrons in the floating gate make the corresponding cell's threshold voltage higher.

As shown in Fig.~\ref{fig:structure}, each flash memory block is a two-dimensional cell array where each cell is connected to a wordline (WL) and a bitline (BL). For $B$-bit per cell flash memory, each WL stores $B$-page data.

\begin{figure}[!t]
\centering
\subfloat[Single-level cell (SLC) for $B=1$]{\includegraphics[width=3in]{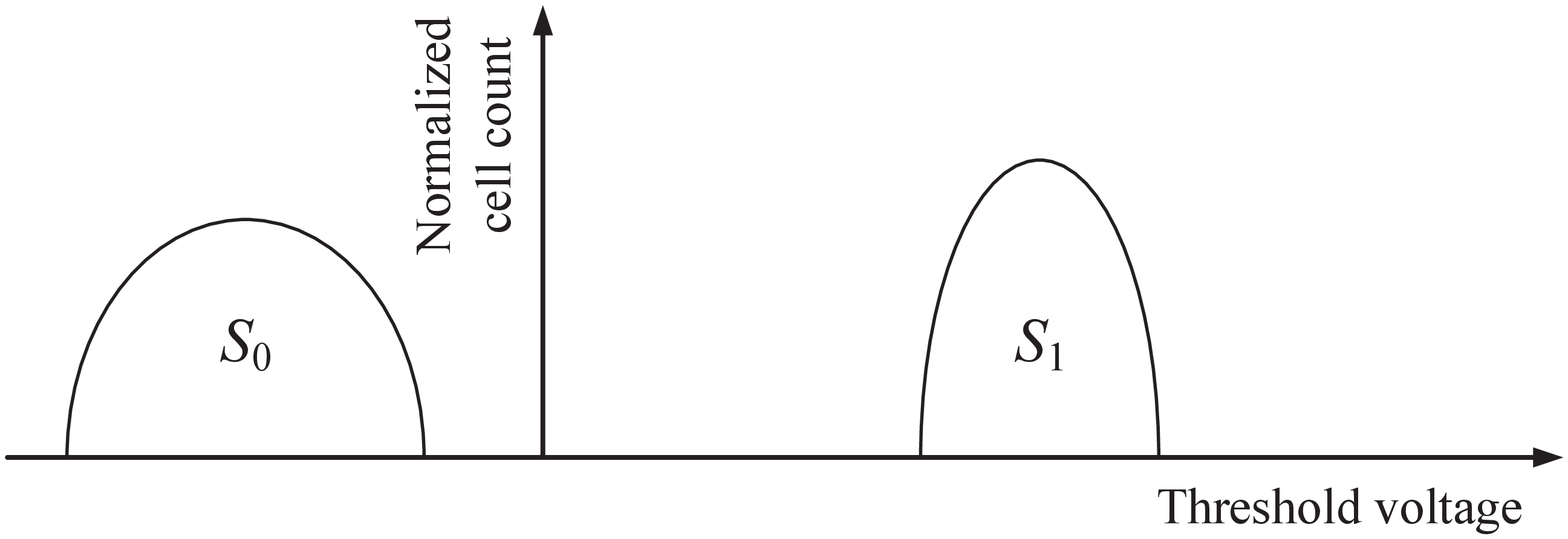}
\label{fig:slc}}
\hfil
\vspace{-2mm}
\subfloat[Multi-level cell (MLC) for $B=2$]{\includegraphics[width=3in]{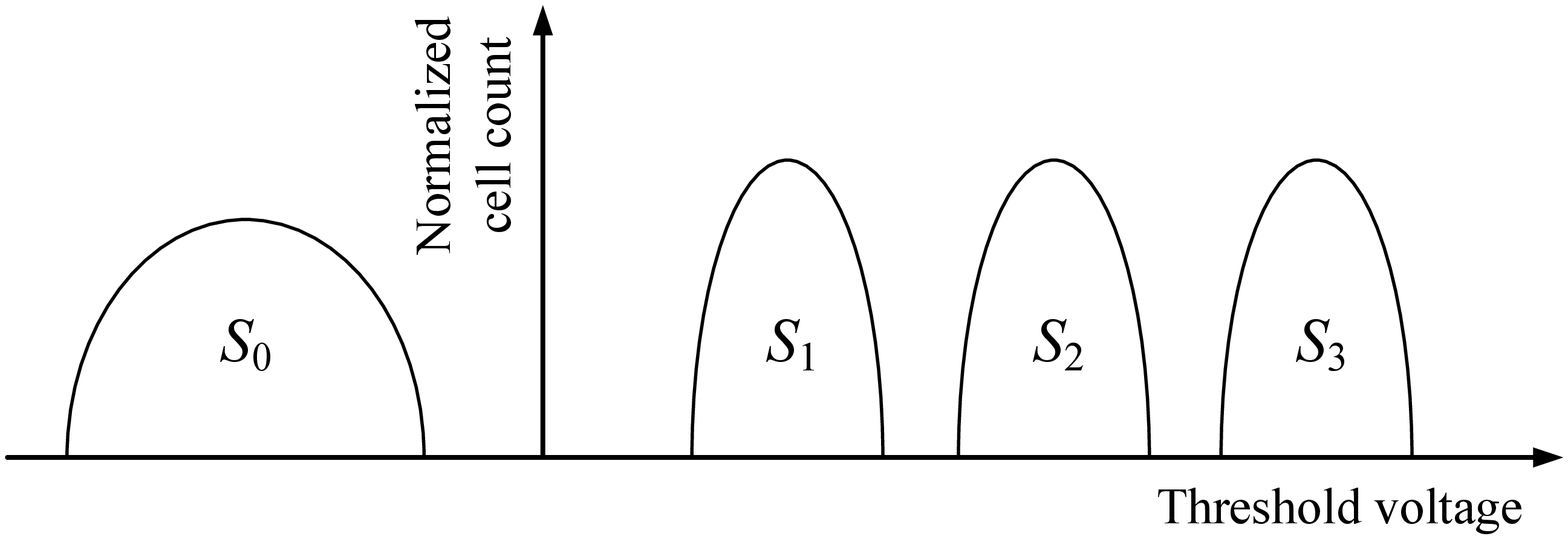}
\label{fig:mlc}}
\caption{Threshold voltage distribution of flash memory cells.}
\label{fig:slc_mlc}
\vspace{-5mm}
\end{figure}

In order to store $B$-bit per cell, each cell's threshold voltage is divided into $2^B$ states, which is similar to pulse amplitude modulation (PAM). Fig.~\ref{fig:slc_mlc}~\subref{fig:slc} shows the threshold voltage distribution of 1-bit per cell flash memory, which is traditionally called single-level cell (SLC). Initially, all memory cells are erased, so their threshold voltage is in the lowest state $S_0$. In order to store data, some of cells in $S_0$ should be written (programmed) into $S_1$. For multi-level cell (MLC) flash memories (i.e., $B \ge 2$), some of cells in $S_0$ (erase state) will be written into $S_1, \ldots, S_{2^B - 1}$ (program states) as shown in Fig.~\ref{fig:slc_mlc}~\subref{fig:mlc}.

The most widely used write operation scheme is the incremental step pulse programming (ISPP) scheme, which was proposed to maintain a tight threshold voltage distribution for high reliability \cite{Suh1995}. The ISPP is based on repeated program and verify cycles with the staircase program voltage $V_{\textrm{pp}}$. Each program state associates with a verify level that is used in the verify operation. During each program and verify cycle, the floating gate threshold voltage is boosted by up to the incremental step voltage $\Delta V_{\textrm{pp}}$ and then compared with the corresponding verify level. If the threshold voltage of the memory cell is still lower than the verify level, the program and verify iteration continues. Otherwise, further programming of this cell is disabled \cite{Suh1995, Dong2010compensation}.

The positions of program states are determined by verify levels and the tightness of each program state depends on the incremental step voltage $\Delta V_{\textrm{pp}}$. By reducing $\Delta V_{\textrm{pp}}$, the threshold voltage distribution can be made tighter, however the write time increases \cite{Suh1995, Kim2012verify}.

In read operation, the threshold voltages of cells in the same WL are compared to a given read level. After a read operation, a page of binary data is transferred to the page buffer in Fig.~\ref{fig:structure}. The binary data shows whether the threshold voltage of each cell is lower or higher than the given read level. Namely, the read operation of flash memory is a binary decision. Thus, multiple read operations are required to obtain a soft decision value, which lowers the read speed. The degradation of read speed is an important challenge for soft decision decoding and signal processing \cite{Dong2011soft, Dong2010compensation}.

The threshold voltage of flash memory cell can be reduced by erase operation. In flash memory, all the memory cells in the same flash memory block should be erased at the same time \cite{Suh1995}. Note that a page of data (within a WL) can be written or read (generally, a flash memory block consists of 64 WLs \cite{Li2009}). In addition, the threshold voltage of cell should be moved into the lowest state $S_0$ by erase operation whereas a slight increase of threshold voltage is possible by ISPP during write operation \cite{Suh1995}. These unique properties of flash memory cause \emph{asymmetry between write and erase operations}.

\subsection{Inter-cell Interference}

\begin{figure}[t]
   \centering
   \includegraphics[width=0.37\textwidth]{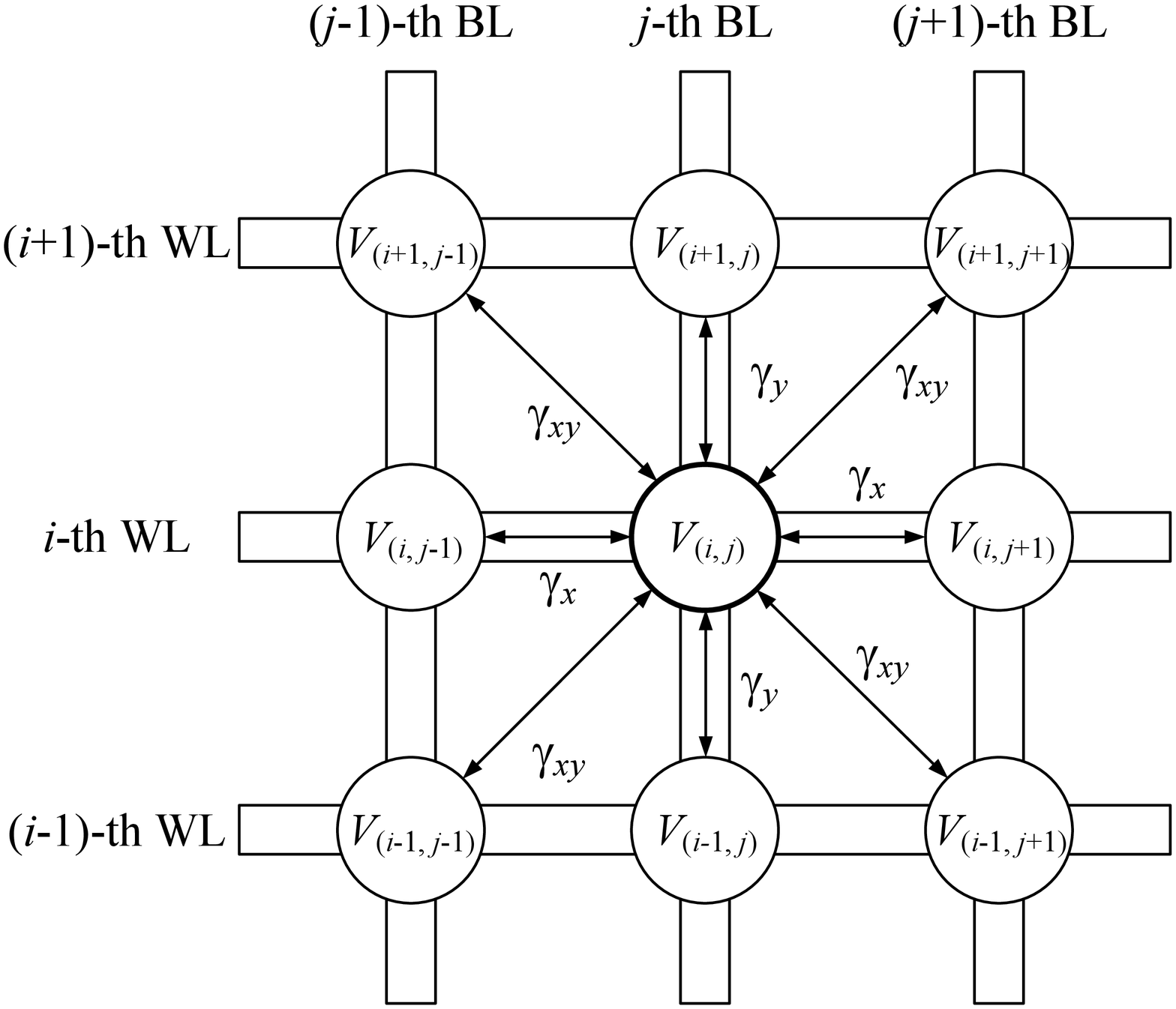}
   \vspace{-3mm}
   \caption{Inter-cell interference between adjacent cells.}
   \label{fig:ici}
   \vspace{-4mm}
\end{figure}

In flash memory, the threshold voltage shift of one cell affects the threshold voltage of its adjacent cell because of the ICI. The ICI is mainly attributed to parasitic capacitances coupling effect between adjacent cells \cite{Lee2002, Prall2007}.

Fig.~\ref{fig:ici} illustrates the ICI between adjacent cells. $V_{(i,j)}$ is the threshold voltage of $(i, j)$ cell which is situated at $i$-th WL and $j$-th BL. $\gamma_x$ is $x$-directional coupling ratio between BL and adjacent BL. Also, $\gamma_y$ is $y$-directional coupling ratio between WL and adjacent WL. Finally, $\gamma_{xy}$ is
$xy$-directional (diagonal) coupling ratio. These coupling ratios depend on parasitic capacitances between adjacent cells. As the cell size continues to shrink, the distances between cells become smaller, which results in the increase of the parasitic capacitances. The increase of parasitic capacitances causes the increase of coupling ratios \cite{Lee2002, Prall2007}.

According to \cite{Lee2002}, the threshold voltage shift $\Delta_{\textrm{ICI}}V_{(i, j)}$ of $(i,j)$ cell due to the ICI is given by
\begin{equation} \label{eq:ici}
\begin{aligned}
\Delta_{\textrm{ICI}}V_{(i, j)} &= \gamma_x \left( \Delta V_{(i, j-1)} + \Delta V_{(i, j+1)} \right) \\
    &+ \gamma_y \left( \Delta V_{(i-1, j)} + \Delta V_{(i+1, j)} \right) \\
    &+ \gamma_{xy} \left( \Delta V_{(i-1, j-1)} + \Delta V_{(i-1, j+1)} + \right.\\
    &\quad\quad\quad \left. \Delta V_{(i+1, j-1)} + \Delta V_{(i+1, j+1)} \right)
\end{aligned}
\end{equation}
where $\Delta V_{\left(i\pm1, j\pm1\right)}$ in the right hand side represent the threshold voltage shifts of adjacent cells after the $(i,j)$ cell has been written. The ICI that happens before writing  $(i,j)$ cell can be compensated by several write schemes so long as $(i,j)$ cell is in program states \cite{Park2008zeroing, Shibata2008}. Note that the ICI to $(i,j)$ cell in $S_0$ cannot be compensated by these write schemes since a cell in $S_0$ is never written (i.e., stay in $S_0$) \cite{Shibata2008, Kim2013modulation}.

\section{Writing on Dirty Flash Memory}\label{sec:dirtyflash}

\subsection{Flash Memory Channel}\label{subsec:fmc}

The flash memory channel can be given by
\begin{align}
Y &= X + S + Z \label{eq:fmc0} \\
  &= X + Z_\textrm{write} + S + Z_\textrm{read} \label{eq:fmc1} \\
  &= V + S + Z_\textrm{read} \label{eq:fmc2}
\end{align}
where $X$ and $Y$ are the channel input and output. Also, $S$ represents the ICI from adjacent cells. The additive random noise $Z$ is a sum of $Z_\textrm{write}$ and $Z_\textrm{read}$ where $Z_\textrm{write}$ is the write noise due to the initial threshold voltage distribution after erase operation and the incremental step voltage $\Delta V_{\textrm{pp}}$ of ISPP. $Z_\textrm{read}$ is the read noise due to other noise sources.

Since the write noise $Z_\textrm{write}$ precedes the ICI $S$, we consider a random variable $V = X + Z_\textrm{write}$. As shown in \eqref{eq:ici}, the shifts of $V$ in adjacent cells determine the ICI $S$. Thus, we claim that the ICI $S$ of $(i, j)$ cell is the same as $\Delta_{\textrm{ICI}}V_{(i, j)}$ of \eqref{eq:ici}. The read noise $Z_\textrm{read}$ happens after ICI. The channel model of \eqref{eq:fmc0} was validated by the real data from the 2x nm NAND flash memory \cite{Moon2013}.

It seems that \eqref{eq:fmc0} is the same as \eqref{eq:costa}. However, there are important differences between the flash memory channel and the dirty paper channel of \eqref{eq:costa}. First, $S$ of \eqref{eq:fmc0} depends on the adjacent cells' $X$ and $Z_\text{write}$ whereas $S$ of \eqref{eq:costa} are independent. In addition, the mean of $S$ in \eqref{eq:fmc0} is not zero since the coupling ratios $\left(\gamma_x, \gamma_y, \gamma_{xy} \right)$ are positive and the threshold voltage shifts of adjacent cells $\Delta V_{\left(i\pm1, j\pm1\right)}$ in \eqref{eq:ici} are nonnegative.

\begin{figure}[!t]
\centering
\subfloat[Threshold voltage distribution of cells in the $i$-th WL before writing the $(i-1)$-th WL]{\includegraphics[width=2.3in]{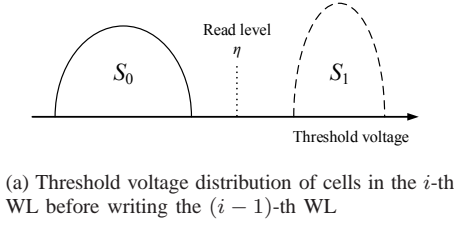}
\label{fig:preread_a}}
\vspace{-3mm}
\hfil
\subfloat[Threshold voltage distribution of cells in the $i$-th WL after writing the $(i-1)$-th WL (still, before writing the $i$-th WL)]{\includegraphics[width=2.3in]{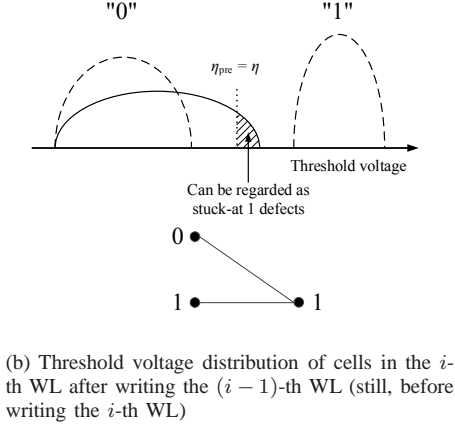}
\label{fig:preread_b}}
\caption{Change from the flash memory channel with the ICI to the model of memory with defective cells by one pre-read operation.}
\label{fig:preread}
\vspace{-5mm}
\end{figure}

Now, we discuss why it is difficult for the encoder to know $S$. First, the encoder has to know the channel input $X$ of adjacent cells in different WLs. Since the write operation is performed page by page, it is possible for the encoder to know the channel input of cells in several WLs only the case where a large number of continuous pages are written at a time \cite{Dong2010compensation}.

Even in the case where the encoder knows enough channel input $X$ of several WLs in advance, it is still difficult to know the random variable $V = X + Z_\textrm{write}$ that determines the ICI. Note that the write noise $Z_\textrm{write}$ is random. In addition, it is much more complicated to know the voltage shift of adjacent cells (i.e., $\Delta V_{\left(i\pm1, j\pm1\right)}$ in \eqref{eq:ici}) since flash memory's read operation is inherently binary decision. Thus, multiple read operations are required to know $\Delta V_{\left(i\pm1, j\pm1\right)}$.


\subsection{Dirty Flash Memory}\label{subsec:dfm}

We describe how to change the flash memory channel with the ICI into the model of memory with defective cells. By this change, the encoder can readily obtain the side information of defects $S^+$ instead of the side information of ICI $S$.

Consider an SLC flash memory. Fig.~\ref{fig:preread} shows the threshold voltage distribution of cells in the $i$-th WL before writing. Initially, all cells are in the erase state $S_0$ as shown in Fig.~\ref{fig:preread}~\subref{fig:preread_a}. However, after writing the adjacent $(i-1)$-th WL, the threshold voltages of cells in the $i$-th WL will be distorted due to the ICI from the $(i-1)$-th WL as shown in Fig.~\ref{fig:preread}~\subref{fig:preread_b}. Thus, some of cells' threshold voltages can be higher than the given read level $\eta$ though the $i$-th WL has yet to be written.

\begin{figure}[t]
   \centering
   \includegraphics[width=2.3in]{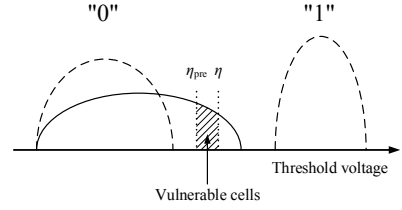}
   \caption{Vulnerable cells can also be regarded as stuck-at 1 defects by setting a pre-read level such that $\eta_\text{pre} < \eta$.}
   \label{fig:preread_different}
   \vspace{-5mm}
\end{figure}

As explained in Section \ref{subsec:basic}, the threshold voltage of flash memory cells cannot be reduced during write operation. In order to decrease the threshold voltage of a cell, we have to erase the whole flash memory block. Thus, the cell whose threshold voltage is higher than the read level $\eta$ will be decided as $S_1$. Assume that $S_0$ and $S_1$ denote the data ``0'' and ``1'' respectively. If a ``0'' is attempted to be written to this cell, an error results. However, ``1'' can be written into this cell. Thus, these cells can be regarded as stuck-at 1 defects in Fig.~\ref{fig:channel_defect}. Even if some of the cells which are regarded as stuck-at 1 defects may be ``0'' due to the read noise $Z_\text{read}$, this error can be corrected by the partitioned linear block codes (PLBC) explained in Section~\ref{subsec:writing_dfm}.

The defect information of stuck-at 1 defects, i.e., $S^+$ can be obtained by just one read operation before writing the $i$-th WL. Before writing the $i$-th WL, the read operation will be performed at the given read level, i.e., \emph{pre-read operation}. When the read level for the pre-read operation, i.e., \emph{pre-read level} $\eta_{\text{pre}}$ is the same as the read level $\eta$, the cells whose threshold voltages are higher than the read level $\eta$ can be identified by the pre-read operation. Thus, the encoder can know the side information of defects $S^+$.

Let $\circ$ denote the operator  $\circ:\{0, 1\} \times \left\{0, 1, \lambda\right\} \rightarrow \{0, 1\}$ by
\begin{equation}\label{eq:circ_operator}
x \circ s^+ =
\begin{cases}
x, & \text{if } s^+ = \lambda ; \\
s^+, & \text{if } s^+ \ne \lambda
\end{cases}
\end{equation}
where $x \in X$ and $s^+ \in S^+$. The binary channel input $X$ and the ternary defect information $S^+$ are shown in Fig.~\ref{fig:channel_defect} \cite{Heegard1983plbc}.

By only one pre-read operation before writing, the flash memory channel with the ICI in \eqref{eq:fmc0} can be changed into the following channel model of \emph{binary} memory with defective cells.
\begin{equation}\label{eq:BDC}
Y = X \circ S^+ + Z
\end{equation}
where $X$, $Y$, and $Z$ are the \emph{binary} channel input, output, and additive noise, respectively. In contrast, $X$, $Y$, and $Z$ of \eqref{eq:fmc0} are \emph{real} values. Note the difference between ``$X+S$'' in \eqref{eq:fmc0} and ``$X \circ S^+$'' in \eqref{eq:BDC}.

It is worth mentioning that $S^+$ does not reveal the $x$-directional ICI from the $i$-th WL and the ICI from the $(i+1)$-th WL, which are subsequent ICI since the pre-read operation is done before the write operations of $i$-th and $(i+1)$-th WLs.

However, the effect of these subsequent ICI can be alleviated by changing the pre-read level. Suppose that the pre-read level $\eta_{\text{pre}}$ is lower than the read level $\eta$ as shown in Fig.~\ref{fig:preread_different}. A cell whose threshold voltage is between $\eta_{\text{pre}}$ and $\eta$ is a vulnerable cell though it is not a stuck-at 1 defect. When the data ``0'' is written to this cell, the ISPP cannot change the threshold voltage of this cell and its threshold voltage is near $\eta$. Thus, it is vulnerable to the subsequent ICI and read noise. On the other hand, the cell's threshold voltage will be higher than a verify level of $S_1$ by the ISPP when the data ``1'' is written to this cell. Note that the verify level of $S_1$ is higher than the read level $\eta$.

Thus, by setting a pre-read level such that $\eta_{\text{pre}} < \eta$, we can regard all the cells whose threshold voltages are higher than the pre-read level $\eta_{\text{pre}}$ as stuck-at 1 defects. Using the additive encoding, only the data ``1'' will be written to these cells. Thus, we can obtain more noise margin between $S_0$ and $S_1$ and prevent the subsequent ICI and read noise.

Imagine a sheet of \emph{lined} paper (Costa considered a sheet of \emph{blank} paper in \cite{Costa1983dpc}). A flash memory block is a sheet of paper and each WL corresponds to a row between lines. If a row between lines is spacious, then the writer can easily write a message between lines. In order to write more messages on a sheet of paper, the writer tries to narrow the space between lines (i.e., scaling down). However, as the space between lines narrows, it is more difficult to write a message without crossing the lines (i.e., ICI). Eventually, after writing a message in a narrower space, the adjacent rows have more dirty spots (i.e., stuck-at 1 defects) due to the ink marks crossing the line. One way to solve this problem is to erase the dirty spots in a corresponding row before writing. However, erasing a row is not permitted (because of the asymmetry between write and erase in flash memory).

Now we consider the other way instead of erasing a row before writing. Assume that the writer knows the location of the dirty spots, but the reader cannot distinguish between the message and the dirt \cite{Costa1983dpc}. Hence, the problem of writing on flash memory with the ICI can be considered as a Costa's writing on dirty paper, i.e., \emph{writing on dirty flash memory}. Since the dirty spots are changed into stuck-at 1 defects by the pre-read operation, writing on dirty flash memory is equivalent to writing on (flash) memory with defective cells. Thus,``writing on dirty paper" in communication and ``memory with defective cells" in storage come together in flash memory.

%

\subsection{Writing on Dirty Flash Memory}\label{subsec:writing_dfm}

Now the encoder knows the side information of stuck-at 1 defects of flash memory, i.e., the writer knows the dirty spots on a sheet of lined paper. The next step is to write a message on dirty flash memory taking into account that the decoder cannot distinguish between the 1s in a message and the stuck-at 1 defects, i.e., the reader cannot distinguish between the ink marks and the dirty spots.

Tsybakov proposed the \emph{additive encoding} approach which masks defects by adding a carefully selected binary vector \cite{Tsybakov1975additive, Tsybakov1975}. Masking defects is to make a codeword whose values at the locations of defects match the stuck-at values at those locations. Heegard elaborated the additive encoding and defined the $[n, k, l]$ partitioned linear block codes (PLBC) that mask stuck-at defects and correct random errors \cite{Heegard1983plbc}.

We will apply the $[n, k, l]$ PLBC for writing on dirty flash memory. A vector version of \eqref{eq:BDC} for an $n$-cell memory is given by
\begin{equation}\label{eq:BDC_vector}
\mathbf{y} = \mathbf{x} \circ \mathbf{s}^+ + \mathbf{z}
\end{equation}
where $\mathbf{x}, \mathbf{y}, \mathbf{z} \in \left\{0, 1\right\}^n$ are the binary channel input vector, output vector, and random error vector, respectively. Also, $\mathbf{s}^+ \in \left(S^+\right)^n$ represents the side information of defect locations and stuck-at values. Both $\circ$ and $+$ are the vector component-wise operators. 

The number of defects in $n$ cells is equal to the number of non-$\lambda$ components in $\mathbf{s}^+$. The number of errors due to defects is given by
\begin{equation}\label{eq:BDC_num_errors}
\| \mathbf{x} \circ \mathbf{s}^+ - \mathbf{x} \|
\end{equation}
where $\| \cdot \|$ is the Hamming weight of the vector.

The $[n,k,l]$ PLBC is a pair of linear subspaces $\mathcal{C}_1 \subset \left\{0, 1\right\}^n$ and $\mathcal{C}_0 \subset \left\{0, 1\right\}^n$ of dimension $k$ and $l$ such that $\mathcal{C}_1 \cap \mathcal{C}_0 =\{ \mathbf{0}\}$. The encoding and decoding of the $[n,k,l]$ PLBC are summarized as follows. (The details were presented in \cite{Heegard1983plbc}.)

\emph{Encoding:} A message $\mathbf{m} \in \left\{0, 1\right\}^k$ is encoded to a corresponding codeword $\mathbf{c} \in \mathcal{C}$ as follows.
\begin{align}
\mathbf{c} &= G_1\mathbf{m} + G_0\mathbf{d} = \left[G_1 \: G_0 \right] \begin{bmatrix} \mathbf{m} \\ \mathbf{d} \end{bmatrix} \label{eq:encoding0}\\
    &= \widetilde{G} \begin{bmatrix} \mathbf{m} \\ \mathbf{d} \end{bmatrix} \label{eq:encoding1}
\end{align}
where $\mathbf{c}_1 = G_1\mathbf{m} \in \mathcal{C}_1$ and $\mathbf{c}_0 = G_0\mathbf{d} \in \mathcal{C}_0$. Note that $\mathbf{d} \in \left\{0, 1\right\}^l$ is the parity for masking defects. The generator matrix $G_1$ is an $n \times k$ matrix and the generator matrix $G_0$ is an $n \times l$ matrix. Thus, $\mathcal{C}$ can be regarded as an $[n, k+l]$ linear block code with the generator matrix $\widetilde{G} = \left[G_1 \: G_0 \right]$.

\emph{Decoding:} Retrieve $\mathbf{y} = \mathbf{x} \circ \mathbf{s}^+ + \mathbf{z}$ where $\mathbf{x} = \mathbf{c}$. Compute the syndrome $\mathbf{w} = \widetilde{H}^T \mathbf{y}$ (superscript $T$ denotes transpose) and choose $\widehat{\mathbf{z}} \in \left\{0, 1\right\}^n$ which minimizes $\| \mathbf{z} \|$ subject to $\widetilde{H}^T \mathbf{z} = \mathbf{w}$. Then $\widehat{\mathbf{m}} =  \widetilde{G}_1^T \widehat{\mathbf{c}}$ where $\widehat{\mathbf{c}}= \mathbf{y} + \widehat{\mathbf{z}}$. The parity check matrix $\widetilde{H}$ is an $n \times r$ matrix such that $\widetilde{H}^T \widetilde{G} = 0_{r, k+l}$ (the $r \times (k+l)$ zero matrix) and $k+l+r=n$. The message inverse matrix $\widetilde{G}̃_1$ is defined as an $n \times k$ matrix such that $\widetilde{G}̃_1^{T} G_1 =I_k$ (the $k$-dimensional identity matrix) and $\widetilde{G}̃_1^{T} G_0 =0_{k, l}$.

The parity $\mathbf{d}$ for masking defects determines the binary vector $\mathbf{c}_0$ masking stuck-at defects. The encoder should choose $\mathbf{d}$ judiciously by considering both $\mathbf{c}_1$ and $\mathbf{s}^+$. The optimal $\mathbf{d}$ is chosen to minimize the number of errors due to defects, i.e. $\| \mathbf{c} \circ \mathbf{s}^+ - \mathbf{c} \|$. Since the computational complexity for finding the optimal $\mathbf{d}$ is exponential, we use the \emph{two-step encoding scheme} for determining $\mathbf{d}$ which was proposed in \cite{Kim2013coding, Kim2013redundancy}.

If $\| \mathbf{c} \circ \mathbf{s}^+ - \mathbf{c} \| \ne 0$, there are errors due to unmasked defects. Since $S^+$ contains partial information of $S$, the remaining ICI which is not included in $S^+$ results in errors. Also, we should consider the random errors in cells (even the cells regarded as stuck-at 1 defects) due to the read noise $Z_\text{read}$ in \eqref{eq:fmc2}. All these errors will be regarded as random errors during decoding.

\section{Numerical Results}\label{sec:result}

\begin{table}[t]
\renewcommand{\arraystretch}{1.3}
\caption{Simulation Parameters}
\label{tab:parameters}
\centering
{\small
\begin{tabular}{|c|c|}
\hline
Parameters & Values   \\ \hline \hline
Bits per cell & $B=1$ (SLC)  \\ \hline
Architecture & All bitline (ABL) \\ \hline
Initial threshold voltage & \multirow{2}{*}{$\mathcal{N} \left(-3, 1^2\right)$}  \\
distribution & \\ \hline
Verify level for $S_1$ & $V_{S_1} = 1$  \\ \hline
Incremental step voltage & $\Delta V_\textrm{pp} = 1$ \\ \hline
Coupling ratios $\left(\gamma_x, \gamma_y, \gamma_{xy} \right)$ & $\alpha \left(0.08, 0.1, 0.006 \right)$ \\ \hline
$Z_\textrm{read}$ of \eqref{eq:fmc2} & $\mathcal{N} \left(0, \sigma_{Z_\text{read}}^2 \right)$ \\ \hline
Read level between $S_0$ and $S_1$ & $\eta = 0$  \\ \hline
\multirow{2}{*}{Additive encoding} & $\left[ n = 1023, k=923, l \right]$ \\
 & PBCH Codes \\ \hline
\end{tabular}}
\end{table}

\begin{table}[t]
\renewcommand{\arraystretch}{1.3}
\caption{All Possible Redundancy Allocation Candidates of $\left[ n = 1023, k=923, l \right]$ PBCH Codes}
\label{tab:PLBC}
\centering
{\small
\begin{tabular}{|c|c|c|c|}
\hline
Code & {$l$} & {$r$} & Notes   \\ \hline \hline
0 & 0 & 100 & Only correcting random errors \\ \hline
1 & 10 & 90 &\\ \hline
2 & 20 & 80 &\\ \hline
3 & 30 & 70 &\\ \hline
4 & 40 & 60 &\\ \hline
5 & 50 & 50 &\\ \hline
6 & 60 & 40 &\\ \hline
7 & 70 & 30 &\\ \hline
8 & 80 & 20 &\\ \hline
9 & 90 & 10 &\\ \hline
10& 100 & 0 & Only masking defects\\ \hline
\end{tabular}}
\vspace{-5mm}
\end{table}

In this section, we present the numerical results when the encoder uses the side information of $\mathbf{s}^+$ in \eqref{eq:BDC_vector}. The simulation parameters are summarized in Table.~\ref{tab:parameters}. The initial threshold voltage distribution (after erasing a flash memory block) is assumed to be the Gaussian distribution $\mathcal{N} \left(-3, 1^2\right)$. The ISPP was implemented with the parameters of the verify level for $S_1$, i.e., $V_{S_1} = 1$ and the incremental step voltage $\Delta V_\textrm{pp} = 1$. Note that the variance of initial threshold voltage distribution and the incremental step voltage work for $Z_\textrm{write}$ of \eqref{eq:fmc1}, which precedes the ICI.

The ICI $S$ is calculated by \eqref{eq:ici} where the coupling ratios are given by $\left(\gamma_x, \gamma_y, \gamma_{xy} \right) = \alpha \left(0.08, 0.1, 0.006 \right)$. The scaling factor $\alpha$ represents the ICI strength, and the ratios between $\gamma_x$, $\gamma_y$, and $\gamma_{xy}$ are taken from \cite{Prall2007}. These ratios can be different for each product of flash memory. The read noise $Z_\textrm{read}$ after the ICI is assumed to the $\mathcal{N} \left(0, \sigma_{Z_\textrm{read}}^2 \right)$.

For additive encoding, we consider $\left[ n = 1023, k=923, l \right]$ partitioned Bose, Chaudhuri, Hocquenghem (PBCH) codes. The PBCH code is a special class of PLBC, which can be designed by a similar method of standard BCH codes \cite{Heegard1983plbc}. For the given $n=1023$ and $k=923$, all possible redundancy allocation candidates of PBCH codes are presented in Table~\ref{tab:PLBC}. 

\begin{figure}[t]
   \centering
   \includegraphics[width=3in]{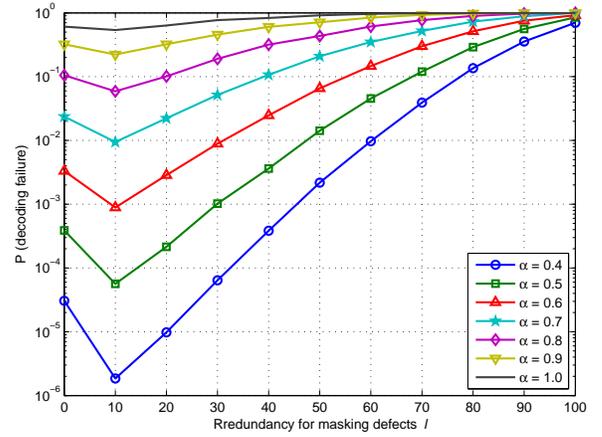}
   \caption{Comparison of $P(\text{decoding failure})$ for different scaling factors $\alpha$ ($\sigma_{{Z_\textrm{read}}} = 0.1$, $\eta_\text{pre} = \eta = 0$).}
   \label{fig:plot_wer}
  \vspace{-5mm}
\end{figure}

\begin{figure}[!t]
\centering
\subfloat[$\sigma_{{Z_\textrm{read}}} = 0.1$]{\includegraphics[width=3in]{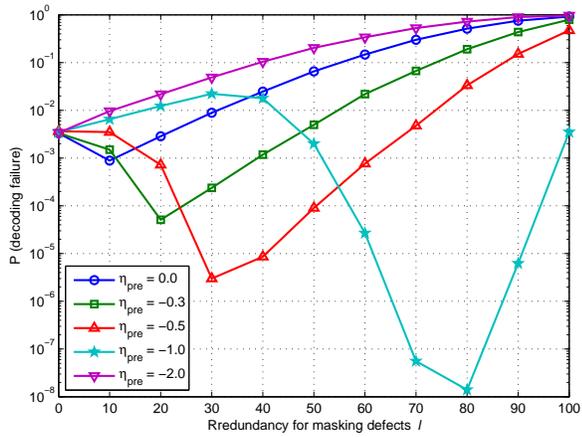}
\label{fig:plot_wer_preread_a}}
\hfil

\subfloat[$\sigma_{{Z_\textrm{read}}} = 0.3$]{\includegraphics[width=3in]{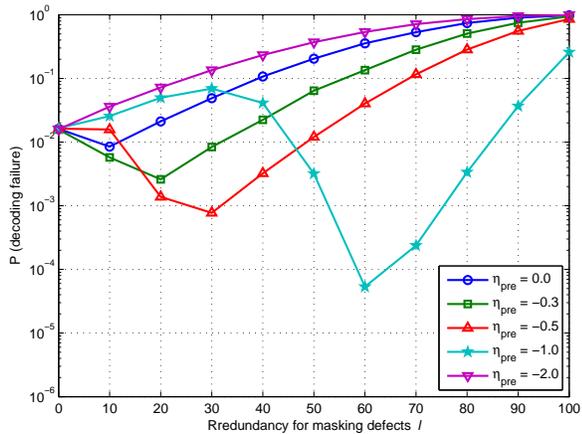}
\label{fig:plot_wer_preread_b}}

\caption{Comparison of $P(\text{decoding failure})$ for different pre-read levels ($\alpha = 0.6, \eta_\text{pre} \le \eta$).}
\label{fig:plot_wer_preread}
\vspace{-5mm}
\end{figure}

%

%

Fig.~\ref{fig:plot_wer} shows that $P(\text{decoding failure})$ can be improved by using the side information of $\mathbf{s}^+$. If the redundancy for masking $l$ is zero, it means that the side information is ignored. Otherwise, the encoder uses the side information and can improve $P(\text{decoding failure})$. The optimal redundancy allocation $(l, r)$ to minimize $P(\text{decoding failure})$ depends on the ICI $S$ and the additive noise $Z$. For the given parameters, the optimal redundancy allocation is $(l, r) = (10, 90)$. The pre-read level $\eta_\text{pre}$ is set to zero ($\eta_\text{pre} = \eta = 0$).

\begin{figure}[!t]
\centering
\includegraphics[width=3in]{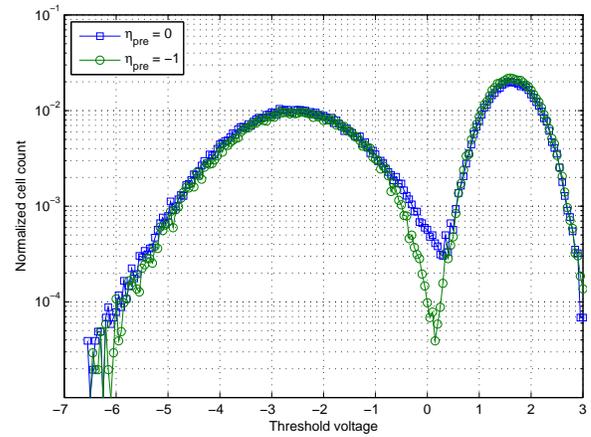}
\caption{Comparison of threshold voltage distributions for different pre-read levels $\eta_\text{pre}$ $\left(\alpha = 0.6, \sigma_{{Z_\textrm{read}}} = 0.3, (l, r) = (100, 0), \eta=0 \right)$.}
\label{fig:plot_distribution}
\vspace{-5mm}
\end{figure}

Fig.~\ref{fig:plot_wer_preread} shows that changing the pre-read level $\eta_\text{pre}$ can improve $P(\text{decoding failure})$ significantly. This improvement can be explained by Fig.~\ref{fig:plot_distribution}. Compare two threshold voltage distributions of $\eta_\text{pre} = 0$ and $\eta_\text{pre} = -1$. The threshold voltage distribution of $\eta_\text{pre} = -1$ is better than that of $\eta_\text{pre} = 0$. The cells whose threshold voltages are between $\eta_\text{pre}$ and $\eta$ are vulnerable to the subsequent ICI and read noise since these cells' threshold voltages are highly probable to be higher than $\eta$. By setting $\eta_\text{pre} < \eta$, we can also regard these vulnerable cells as stuck-at 1 defects. Thus, all the cells whose threshold voltages are higher than $\eta_\text{pre}$ will be written into $S_1$, which results in the improvement of the threshold voltage distributions as shown in~Fig.~\ref{fig:plot_distribution}.

As the gap between $\eta_\text{pre}$ and $\eta$ is bigger, more cells are regarded as stuck-at 1 defects. More defects require more redundancy $l$ for masking defects. If the number of defects is too large, the additive encoding fails to mask defects sufficiently. Thus, $P(\text{decoding failure})$ of $\eta_\text{pre}=-2$ is worse than that of $P(\text{decoding failure})$ of $\eta_\text{pre}=0$ as shown in Fig.~\ref{fig:plot_wer_preread}.

By comparing Fig.~\ref{fig:plot_wer_preread}~\subref{fig:plot_wer_preread_a} and Fig.~\ref{fig:plot_wer_preread}~\subref{fig:plot_wer_preread_b}, the effect of the read noise can be explained. When $\eta_{\text{pre}} = -1.0$, the optimal redundancy allocation $(l, r)$ in Fig.~\ref{fig:plot_wer_preread}~\subref{fig:plot_wer_preread_a} (for $\sigma_{{Z_\textrm{read}}} = 0.1$) is $(l, r) = (80, 20)$ and the optimal redundancy allocation
in Fig.~\ref{fig:plot_wer_preread}~\subref{fig:plot_wer_preread_b} (for $\sigma_{{Z_\textrm{read}}} = 0.3$) is $(l, r) = (60, 40)$. It is because more redundancy $r$ for correcting random errors should be allotted as the read noise increases.

\section{Conclusion}\label{sec:conclusion}

In this paper, the famous examples of Gelfand-Pinsker problem such as ``writing on dirty paper'' and ``memory with defective cells'' come together in flash memory. The flash memory channel with the ICI which is similar to the channel of writing on dirty paper has been changed into the model of memory with defective cells by only one pre-read operation. The unique properties of asymmetry between write and erase operations and ICI play a pivotal role in this change. When a message is written on dirty flash memory, the dirty spots due to the ICI can be hidden to the reader. Although this paper focused on SLC flash memory, the proposed scheme can be extended to MLC flash memory.



%


\IEEEtriggeratref{20}
\bibliographystyle{IEEEtran}
\bibliography{IEEEabrv,dirty_flash}

\end{document}